\documentstyle[epsfig,epsf]{EuroPhys}

\newif\ifboo \boofalse

\begin{document}
\euro{}{}{}{}
\Date{}
\shorttitle{Mesoscopic Quantum Eraser}
\title{A Mesoscopic Quantum Eraser}
\author{G. Hackenbroich \inst{1}, B. Rosenow \inst{2}, and 
H.A. Weidenm\"uller \inst{2}}
\institute{\inst{1} Universit\"at GH Essen, Fachbereich 7, D--45117
  Essen,
Germany\\
\inst{2} MPI f\"ur Kernphysik, D--69029 Heidelberg,  
Germany} 
\rec{}{}
\pacs{
\Pacs{03}{65.Bz}{Foundations,theory of measurement}
\Pacs{73}{23.-b}{Mesoscopic systems}
\Pacs{73}{50.Bk}{General theory, scattering mechanisms}}
\maketitle
\begin{abstract}
  Motivated by a recent experiment by Buks et al.\ [Nature {\bf 391}, 871
  (1998)] we consider electron transport through an Aharonov--Bohm
  interferometer with a quantum dot in one of its arms.  The quantum dot is
  coupled to a quantum system with a finite number of states acting as a
  which--path detector. The Aharonov--Bohm interference is calculated using a 
  two--particle
  scattering approach for the joint transitions in detector and quantum dot.
  Tracing over the detector yields dephasing and a reduction of the
  interference amplitude. We show that the interference can be restored by a
  suitable measurement on the detector and propose a mesoscopic quantum eraser
  based on this principle.
\end{abstract}

Recent progress in quantum and atom optics has made it possible to
test basic tenets of quantum physics. Complementarity has been tested
in various realizations \cite{zeili} of the classical double--slit
gedanken experiment using photon pairs created in parametric 
down--conversion. Related experiments \cite{Rempe} utilizing atomic beams
instead of photons have been performed very recently. These
experiments not only confirmed the destruction of multiple--path
interference due to a which--path measurement. More importantly, they
also demonstrated that the loss of interference need not be
irreversible if the which-path detector is itself a quantum system.
In fact, realizing Scully's \cite{Scully} idea of a quantum eraser it
was shown that the interference can be restored by erasing the
which-path information from the detector in a subsequent measurement.
Quantum detectors which have been used in practical implementation of
quantum erasers include the photon polarization and internal degrees
of freedom of atoms in an atomic beam.

In this Letter, we address the measurement process and the concept of
a quantum eraser in the domain of mesoscopic physics. Specifically, we
propose a semiconductor microstructure which can act as a quantum
eraser. Far from only duplicating and/or corroborating results in
optics, such a device would be of considerable interest in its own
right. First, it would test quantum physics in the domain of solid
state physics. Second, and in contrast to quantum optics, mesoscopic
probes are inevitably coupled to macroscopic bodies (leads etc.),
opening the possibility to address quantitatively the issue of quantum
decoherence. Finally, mesoscopic probes offer the possibility of
practical applications.

Our work is motivated by the first demonstration of controlled
dephasing in a which--path semiconductor device by Buks {\em et al.}
\cite{Buks}. In a pioneering experiment, the authors used an
Aharonov-Bohm (AB) interferometer with a quantum dot (QD) embedded in
one of the arms. The coherent transmission of electrons through the QD
was detected by the capacitive change of the transmission of a quantum
point contact (QPC) located in the immediate vicinity of the dot.
Theoretical work on this experiment employs a rate equation formalism
\cite{Gurvitz}, a change of the quantum state of the environment of
the dot \cite{Buks}, dephasing of the electron on the dot by the
environment \cite{Levi97}, and creation of both virtual and real
excitations in the QPC \cite{AlWiMe97}.

In the experiment by Buks {\em et al.}, the QPC serves as a dephasing device
and tests complementarity. Theoretically, this situation is described by
accounting for the coupling between QD and QPC, and by tracing
over the states of the QPC. However, while this procedure generates dephasing
in the AB--ring it does not provide any information about the actual path of
the electrons through the AB--ring.  To obtain such information an additional
measurement must be performed on the quantum detector coupled to the
AB--ring.  It is the purpose of the present paper to investigate the influence
of such a {\em true which--path measurement} on the interference contrast.  We
address this question by coupling an AB--ring to a quantum detector with a {\em
  finite number of states} (see Fig.~1)). Such a discrete quantum detector
can be realized e.g.\ by utilizing the spin of the electron passing through
the QD, or by a pair of quantum dots coupled capacitively to the QD in the AB
interferometer. An electron passing through the AB interferometer leaves 
which--path information in the detector by changing its quantum state. We
derive the AB--contrast using a novel two--particle scattering approach for joint
transitions through the AB--ring and in the detector. Analyzing various
possible measurements we show that the detector with the AB--ring can be used
as a quantum eraser, i.e., a setup which allows to erase part or all of the
which--path information stored in the detector.

\begin{figure}
\hspace*{2cm}
\epsfig{file=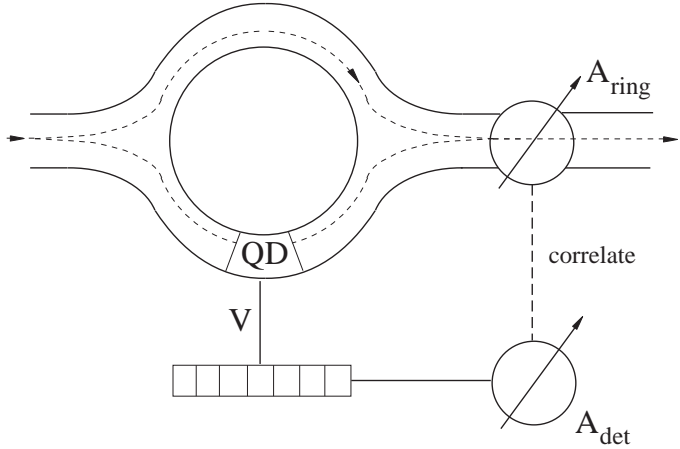,width=9cm,angle=0}
\caption{Schematic view of the mesoscopic quantum eraser. An Aharonov--Bohm 
  interferometer with a quantum dot is coupled to a discrete quantum detector
  via the interaction $V$. The measurement of the electron transmission
  $A_{\rm ring}$ through the interferometer is correlated with the measurement
  $A_{\rm det}$ performed on the detector.}
\label{fig1}
\end{figure}

We assume that the QD is in the Coulomb blockade regime and
sufficiently close to a resonance so that only a single dot state need
be considered.  Energy and width of this resonance in the absence of
any coupling to the quantum detector are denoted by $E_0$ and
$\Gamma$, respectively. Let $\alpha^{},\alpha^\prime$ denote the leads
coupled to the AB interferometer and $p^{},p^\prime$ the transverse
modes in the leads.  Then, $c \equiv \{p \alpha \}$ denotes the
channels with $\delta_{c^{}c^\prime} = \delta_{p^{}p^\prime}
\delta_{\alpha^{} \alpha^\prime}$. We recall that without coupling to
the quantum detector, the scattering matrix $S^{\rm
  AB}_{c^{}c^\prime}$ for passage of an electron through the 
AB--interferometer is given by
\cite{HW}
\begin{equation}
S^{\rm AB}_{c c^\prime}  = \left( S^{(0)}_{c c^\prime} -2
\pi i { \gamma_{c^{}} \gamma_{c^\prime} \over E \! - \! E_0  \! + \!
  i\Gamma /2} \right) \ .
\label{scatt1}
\end{equation}
We have introduced the total energy $E$, and the partial width
amplitudes $\gamma_{c^{}}$, $\gamma_{c^\prime}$ for decay of the QD
resonance into the channels $c^{}$, $c^{\prime}$, respectively. The
term $S^{(0)}$ describes an energy-- and flux--independent background
due to scattering through that arm of the AB--interferometer which does
not contain the QD. The last term in Eq.~(\ref{scatt1}) accounts for
all scattering processes through the QD including multiple scattering
through the AB--ring. This term depends on the magnetic flux $\Phi$
threading the AB--ring through the partial width amplitudes
$\gamma_{c^{}}$, $\gamma_{c^\prime}$ and the total width $\Gamma$
\cite{flux}. The flux--dependence is non--trivial in general since it
includes all harmonics in $\Phi$.

The quantum detector is taken to be an N--state quantum system where
$N = 2$ in the case of the electron spin.  The states of the detector
are labeled $k^{},k^\prime$ with $k^{},k^\prime = 1,\ldots,N$ and
have energies $\epsilon_k$. In order to model the coupling to a double
dot system, or a magnetic field acting on the electron spin and
confined to the QD, we assume that the interaction $V$ between
electron and detector vanishes unless the electron is located on the
QD. With $c^{\dagger}_k$ and $b^{\dagger}$ the creation operators for
the states in the detector and for the state on the QD, respectively,
we account for the coupling to the
detector by adding to the Hamiltonian of Refs.~\cite{HW} the term
\begin{eqnarray} 
H'  = \sum_{k=1}^N \epsilon_k^{} c_k^\dagger c_k^{} +
\sum_{k,k^\prime=1}^N V_{k k^\prime} b^\dagger b^{}
c_k^\dagger c_{k^\prime}^{} \ .
\label{Ham}
\end{eqnarray}

When the coupling $V$ between the interferometer and the detector is
switched on, the quantum states of electron and detector become
entangled and the N--state system can act as a which--path detector.
Electrons going through the quantum dot leave their trace in the
detector by changing its quantum state. We describe this process by
the two--particle scattering amplitude $S_{c c^{\prime} k k^{\prime}}$
for a transition between states $c$ and $c^{\prime}$ in the leads and
the states $k$ and $k^{\prime}$ in the detector. We have derived this
S--matrix using two different methods: (i) by solving the
Lippmann--Schwinger equation and (ii) by using the LSZ--formalism
\cite{LSZ}. The second method allows to us to include screening
effects that arise when the detector itself is a many--body system.
Such effects are e.g.\ important for a QPC-detector \cite{AlWiMe97},
however, they can be neglected for the discrete quantum detectors
considered in the following. Neglecting screening both methods (i), 
(ii) give the result
\begin{equation}
  S_{c c^\prime, k k^\prime} = S^{(0)}_{c c^\prime} \delta_{k
    k^\prime} \! \!- \! 2 \pi i \gamma_c \gamma_{c^\prime} G_{k
    k^\prime} \! ,
\label{scatt2}
\end{equation}
where $S^{(0)}_{c c^\prime}$ was introduced in Eq.\ (\ref{scatt1}).
This expression differs in two important respects from
Eq.~(\ref{scatt1}) obtained for the uncoupled case. First, the
scattering matrix (\ref{scatt2}) allows for energy exchange between
the AB interferometer and the detector. Indeed, in the derivation
there occurs a delta function $\delta_{\epsilon_{c^{}} \! + \!
  \epsilon_{k^{}}, \epsilon_{c^\prime} + \epsilon_{k^\prime}}$ which
expresses the conservation of total energy, while Eq.~(\ref{scatt1})
applies under the condition $\epsilon_{c^{}} = \epsilon_{c^{\prime}}$.
Second, the single Breit--Wigner resonance found in the
non--interacting case has been replaced by the two--particle Green
function $G$ for joint transitions through the dot and in the
detector. The explicit form of $G$ can be found from the matrix
elements
\begin{equation}
[G^{-1}]_{k k^\prime} =  (E - \epsilon_k -E_0+ i \Gamma /2) 
\delta_{k k^\prime}-V_{k k^\prime}
\label{Green}
\end{equation}
of its inverse. We note that the energy $E$ may differ by a constant
from that used in Eq.~(\ref{scatt1}), and that for $V = 0$, $G$ simply
reduces to the product of an $N \times N$ unit matrix and the
Breit--Wigner resonance on the QD. A non--vanishing $V$ leads to a
{\em splitting of the single Breit--Wigner resonance into
  (generically) $N$ resonances}. The positions of these resonances are
determined by the interaction and can be found by diagonalizing the
matrix $\epsilon_k \delta_{k^{}k^\prime} + V_{k^{\prime}k}$. We point
out that the widths of the resulting $N$ resonances are totally
unaffected by the interaction $V$ and given by the width $\Gamma$ of
the uncoupled Breit--Wigner resonance. Hence, the coupling to an
external detector leads to a splitting rather than to a broadening of
the resonance. After taking the trace over the degrees of freedom of
the quantum detector, this splitting can be interpreted as an
effective broadening of the resonance width, see Ref.~\cite{AlWiMe97}.

How does our formalism relate to experiments on the passage of
electrons through the AB interferometer and the interaction with the
quantum detector? Let $\rho^{{(0)}}$ be the density matrix of the
total system prior to the passage of the electron through the AB
interferometer so that $\rho^{{(0)}}$ projects onto states in the lead
{\em feeding} the AB device. 
Let $A=A_{ring} A_{det}$ be the operator of an observable connected to
electron transmission so that $A_{ring}$ projects onto states in the
lead {\em depleting} electrons from the AB device. The expectation
value of $A$ is given by $\langle A \rangle = {\rm Tr} [ \rho A]$. The
trace is taken over the states of detector and leads, and $\rho = S
\rho^{(0)} S^{\dagger}$ is the density matrix after passage through
the AB--device. In taking the traces we treat this final measurement
as an orthodox one. We are permitted to do so if the entangled
electron--detector state maintains coherence until this final
measurement. In discussing possible realizations we will argue that
this is realistic. We do not include a microscopic analysis of this 
final measurement here as the emergence of classical properties from
the microstate of quantum systems is a well--understood process (see,
e.g., Ref.~\cite{om94}).\\
We now focus attention on the interference contribution. To
leading order in the exponentially small transmission through the QD,
this interference contribution is obtained by keeping in the
scattering matrix of Eq.~(\ref{scatt2}) only the lowest harmonic in
the flux $\Phi$. In this approximation, the partial width amplitudes
take on the explicit flux--dependence $\gamma_{c^{}} \gamma_{c^\prime}
\to \exp[i \alpha] \gamma_{c^{}}(0) \gamma_{c^\prime} (0)$ where
$\gamma_{c^{}}(0)$ and $\gamma_{c^\prime}(0)$ are flux--independent
and where $\alpha =2 \pi \Phi/\Phi_0$ with $\Phi_0$ denoting the
elementary flux quantum.  Writing $\rho^{(0)} = \rho^{(0)}_{\rm ring} 
\rho^{(0)}_{\rm det}$, we find
\begin{equation}
\langle A \rangle = A^{(0)}+{\rm Re} \left\{ |t| e^{ i (\alpha-\alpha_0)} 
{\rm Tr_{det}}[  \rho^{(0)}_{\rm det}G \, A_{\rm det}] \right\} + \ldots ,
\label{cont1}
\end{equation}
where the dots indicate small corrections that arise from higher
harmonics in $\Phi$. The term $A^{(0)}$ results from the passage of
the electron through the free arm of the AB--device without QD, and
$|t| e^{i \alpha_0}$ is a non--universal complex amplitude determined
by quantities describing the AB--device. In contradistinction, the
factor 
${\rm Tr_{det}}[  \rho^{(0)}_{\rm det}G \, A_{\rm det}]$
describes the loss of interference induced by the
coupling of the AB--device to the detector. We emphasize that this
factor not only depends on the interaction between the interferometer
and the detector but also on the actual form of the measurement
$A_{\rm det}$ performed on the detector. This important fact
illustrates the fundamental difference between {\em detector--induced
  dephasing} and a true {\em which--path measurement}. Detector--induced
dephasing amounts to putting $A_{\rm det} = 1$ and tracing out the
degrees of freedom of the detector. This procedure reduces the
magnitude of the interference term in Eq.~(\ref{cont1}). In contrast
to other dephasing mechanisms like e.g. electron--electron
interactions, the detector--induced dephasing may be controlled
experimentally. An example is the dephasing by a QPC-detector
\cite{Buks}. By tracing out the detector, however, no information is
obtained about the actual electron path across the AB device. To get
this information one necessarily has to perform a measurement, i.e.,
use $A_{\rm det} \neq 1$. The possible effects of a which--path
measurement are discussed next. 

{\em Quantum eraser.} It is convenient to describe the detector with
$N = 2$ states using a spin--1/2--terminology although our results are
not confined to this case. The coupling of a spin--1/2 to an electron
traveling through an interferometer was also discussed in \cite{Imry}.
We take $\epsilon_1 = \epsilon_2 = 0$ and
choose a basis such that the coupling $V_{k k^\prime}= g \mu B
(\sigma_z)_{k k^\prime}$ is diagonal. Prior to the passage of the
electron, the detector is assumed to be polarized in the +x-direction,
i.e.\ $\rho^{(0)}_{\rm det}=|x,+ \rangle \langle x,+|$.

The basic idea of the quantum eraser is most easily understood in a
wave-function picture \cite{Scully}. In the AB--interferometer, the
amplitude of the incoming spin--polarized electron is split into two
parts $|\Psi_{j} \rangle = |\psi_j \rangle \otimes |x,+ \! \rangle$,
$j = 1,2$, each part passing through one arm of the interferometer.
Passage through the arm containing the QD causes a transition $|x,+\!
\rangle \to |\chi\! \rangle$ in the spin degree of freedom of
$|\Psi_{1} \rangle$, say, while the spin part of $|\Psi_{2} \rangle$
remains unchanged. The total wave function is given by $|\Psi \rangle
= |\Psi_{1} \rangle + |\Psi_{2} \rangle$ and the current measured
after recombining the amplitudes behind the interferometer is computed
from $\langle \Psi | A | \Psi \rangle$. The flux--dependent part is
proportional to $| \langle \chi | A_{\rm det} |x,+ \rangle|$.  Let us
assume for definiteness that in the QD the spin has been rotated by
$\pi$, so that $|\chi \rangle = |x,- \rangle$. If $A_{\rm det}$
projects onto the $\pm x$-direction, the interference signal is
completely wiped out. This fact reflects the complete which--path
information encoded by the spin. If, however, $A_{\rm det}$ measures
the $z$-component of the spin, the which--path  information is irretrievably
lost (erased) and the spin overlap in $| \langle \chi | A_{\rm det}
|x,+ \rangle|$ is finite. In the present example tracing over the
detector variables, i.e., putting $A_{\rm det}=1$, also results in a
vanishing interference term. This shows that dephasing can completely
destroy the interference even if no which--path information was
obtained.

To implement this idea, we consider the contribution of the
interference term to the current through the AB--device. According to
Eq.~(\ref{cont1}), it has the spin dependence 
\begin{equation}
  \Delta I_{AB} \sim \left| <\!x\!,\!+| {\Gamma /2 \over
    E-\!E_0\!  - \Delta \sigma_z /2 + i \Gamma /2} \, A_{\rm
    det}|x\!,\!+\!> \right| \ ,
\label{cont2}
\end{equation}
where $\Delta = 2 g \mu B$ is the resonance splitting. Any significant
change in the spin state requires that the resonance splitting be
larger than the resonance width. Hence we restrict ourselves to the
case $\Delta \gg \Gamma$. In this case, the QD can be operated in two
possible modes depending on the energy $E$ of the incoming electron:
(i) as a device for rotating the spin orientation within the
$x-y$--plane or (ii) as a Stern--Gerlach filter. Case (i) is realized
for $E=E_0$, and the angle of spin precession is $\pi-2\arctan[ \Gamma
/\Delta]$ resulting in a spin state nearly orthogonal to $|x,+ \!
\rangle$. The total current is obtained by tracing out the detector
($A_{\rm det}= 1$) and shows a strongly suppressed interference term
$\Delta I_{AB} \sim \Gamma^{2} /( \Delta^{2}+ \Gamma^{2})$. The
suppression is due to detector--induced dephasing as discussed above. 
Interference can partly be restored by projecting onto the $\pm z$--spin
direction. Then $\Delta I_{AB} \sim (\Gamma /2) / (\Delta^{2} +
\Gamma^{2})^{1/2}$ which amounts to an enhancement of $| \langle \chi
| A_{\rm det} |x,+ \rangle|$ by a factor of order $\Delta /
\Gamma$. We note that the projection onto the $\pm z$--direction does
not correspond to a true which--path measurement since the electrons
transmitted through either arm of the interferometer have a component
in that direction.

In case (ii) we choose $E = E_0 + \Delta /2$ with $\Delta \gg \Gamma$. 
The QD blocks the $(-z)$-component of the spin. Taking the trace over
spin orientations (with \ $A_{\rm det}=1$), one finds that the
interference term has magnitude $\sim 1/2$. By projecting onto the
$(-x)$--direction one performs a true position measurement and only
detects the amplitude passing the QD. The interference term vanishes
completely. A projection onto the vector bisecting between the $x$--
and the $z$--directions reduces the degree of position information
encoded in the spin and improves the contrast as compared to a
spin--independent measurement by a factor $1 + 1 /\sqrt{2}$.

Some comments on the proposed setup are in order. (i) The measurement
performed on the detector does not affect the total current through the
AB--ring. (This would amount to a backward propagation of information in time
and is forbidden). Instead, the simultaneous measurement of electron spin and
interference term picks only electrons with the selected spin orientation and
thus only part of the total current.  (ii) The quantum eraser is not in
disagreement with general principles of dephasing. The entangled system of
electron and quantum detector stays coherent until the final measurement is
performed. The correlated measurement of both electron position and detector
state is formally equivalent to a second interaction between electron and
quantum detector which wipes out part of the decohering effects of the first
one.  In this sense it can be compared to a reabsorption of the excitation the
electron has left in the environment.

{\em Possible realization.} To use the electron spin as a detector in
a mesoscopic quantum eraser, the electron must be prepared in a
spin--polarized state, the spin must be manipulated in one arm of the
AB--interferometer, and its polarization must be measured behind the 
AB--interferometer.  Appropriate experimental techniques in the growing
field of spin--polarized transport have only recently been developed
\cite{Prinz}. With the help of optical techniques, spin--polarized
electrons have been created by circularly polarized laser beams, and
their spin orientation has been analyzed using polarization--resolved
photoluminescence. A different technique uses magnetized ferromagnetic
contacts to inject and detect electrons. Spin polarizations $> 90 \%$
have been achieved \cite{MHGM88}. Experiments have also shown that at low
temperatures the spin polarization can be maintained over distances of
several $10 \mu$ ($\approx 100 \mu$ in Aluminum). To manipulate the
spin, two scenarios are considered. (i) If the material used for the
interferometer has {\em weak} spin--orbit scattering, we consider a
magnetic field $B$ in $z$--direction confined to the QD. 
Such a field can be realized as the fringe field of a microstructured
ferromagnet. One can obtain a field strength of up to one Tesla
spatially confined to a region of less than one $\mu$ \cite{Johnson}.
In this 
case a two--state detector as discussed above is realized. To see this, let the
ground state of the QD filled with $N$ electrons have total spin $S$
and $z$--component $S_z$. We consider the resonant tunneling of an
electron with spin polarization in the $+x$--direction. Adding this
electron to the QD generates two classes of $(N+1)$--electron states
with $S_z^\prime = S_z \pm 1/2$, respectively. (We recall that $| +x
\rangle = ( |+z \rangle + |-z \rangle)/ \sqrt{2}$). At sufficiently
low temperatures, the excited states within each class do not
contribute to transport, and the QD reduces to a two--state system.
The two ground--state energies differ by the Zeeman energy $g \mu B$.
The influence of spin--dependent tunneling into the QD has been
analyzed in \cite{Buettiker}.
(ii) For {\em strong} spin--orbit scattering there exists an intrinsic
energy splitting between spin--up and spin--down electrons even in the
absence of a magnetic field. This effect may be used to realize a
quantum eraser even without the presence of a quantum dot. We recall
the recent idea \cite{DaDa} of using a surface gate on top of one arm
of the interferometer to modulate the surface electric field and,
hence, the strength of the spin--orbit scattering. This setup would
cause a controlled rotation of the electron spin which would implement
the which--path information. Estimates \cite{Prinz,DaDa} for
semiconductor heterostructures with strong spin--orbit scattering
(e.g.\ InGaAs/InAlAs) show that such devices are within the reach of
existing technology.

In summary, we have investigated electron transport through a mesoscopic
interference device that is coupled to a quantum which--path detector. We
demonstrated that this system can be used as a quantum eraser. Possible
realizations using the electron spin as quantum detector are within reach of
present day technology.


\begin{thebibliography}{99}
\bibitem{flux} The flux dependence of $\Gamma$ is of higher order in the
  tunneling matrix elements between the dot and the ring and can generally be
  neglected.
\end{thebibliography}
\end{document}